\documentclass[aps]{revtex4}
\usepackage{graphicx}

\begin{document}

\title[Agglomeration of DNA-looping proteins]{A thermodynamic model
for agglomeration of DNA-looping proteins}

\author{Sumedha}

\affiliation{Institute for Scientific Interchange,  Viale Settimio Severo
65, Villa Gualino, I-10133 Torino, Italy} 

\author{Martin Weigt}

\affiliation{Institute for Scientific Interchange,  Viale Settimio Severo
65, Villa Gualino, I-10133 Torino, Italy}

\begin{abstract}

In this paper, we propose a thermodynamic mechanism for the formation
of transcriptional foci via the joint agglomeration of DNA-looping
proteins and protein-binding domains on DNA: The competition between
the gain in protein-DNA binding free energy and the entropy  loss due
to DNA looping is argued to result in an effective attraction between
loops. A mean-field approximation can be described analytically via a
mapping to a restricted random-graph ensemble having local degree
constraints and global constraints on the number of connected components. 
It shows the emergence of protein clusters containing a finite fraction 
of all looping proteins. If the entropy loss  due to a single DNA loop 
is high enough, this transition is found to be of first order.

\end{abstract}

\maketitle

\section{Introduction}
Understanding the spatial organization of DNA in the cell / the
cellular nucleus and its relation to transcription is  one of the big
challenges in cell biology \cite{Alberts,Cremer,Cook,Kepes,Segre}.  In
this context, the experimental observation of transcription foci is of
great interest: The transcriptional activity is not evenly distributed
inside the cell, but it is concentrated in focal points around
so-called {\it transcription factories} \cite{Cook}. These factories
contain multiple copies of RNA polymerasis, transcription factors and
parts of the machinery for post-transcriptional RNA modifications. In
order to be transcribed, DNA has to loop back to these transcription
factories, it is expected that one factory is surrounded by about
10-20 DNA loops. In this and related phenomenological pictures
\cite{Kepes,Segre} the formation of transcription  factories and DNA
looping are considered to be of fundamental importance for  the
large-scale spatial organization of the transcriptional activity. A
sound theoretical understanding grounded on simple physical mechanisms
is,  however, missing.

The major reason for the increased efficiency of transcription by 
transcription factories is the following: A locally increased 
concentration of transcription factors close to target genes enhances 
recognition of transcription factor binding sites, the volume a 
transcription factor has to search before finding a target gene is
substantially decreased. In bacteria, local concentration effects
are partially achieved by the co-localization of genes coding for
a transcription factor and their target genes along the one-dimensional
chromosome itself \cite{mirny}. By looping, target genes which are
far along the genome can be brought together in cellular / nuclear
space. Very recently, it was shown experimantally \cite{broek} that
compact DNA conformations actually enhance target localization
compared to stretched conformations. This observation supplies strong
support for the importance of coupling transcriptional activity to 
the spatial DNA organization.

The formation of single DNA loops and its consequences for gene
regulation have recently been in the center of interest of many
bio-physical research works.  These range from precise numerical
descriptions of the looping properties of DNA resp. chromatin fibers
\cite{Bon,Toan} up to the thermodynamic modeling of  mechanisms for
transcriptional gene-regulation. Both direct looping by bivalent
transcription factors (as e.g. the lac repressor) \cite{Vilar,Hwa1}
and  looping via attractive protein-protein interactions between
DNA-bound proteins have been studied \cite{Hwa2,Saiz}. The latter
process is important in particular in distal gene regulation in
eukaryotic cells \cite{Alberts}.

\begin{figure}[htb]
\vspace{0cm}
\begin{center}
\includegraphics[width=0.6\columnwidth]{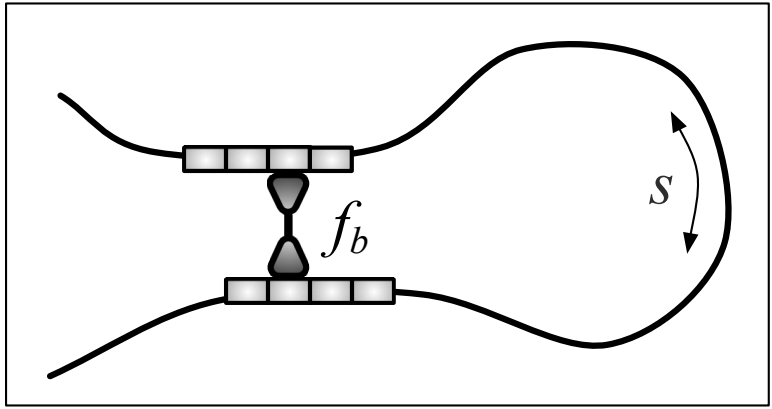}
\end{center}
\caption{Schematic representation of a single DNA loop with one
looping  protein: The looping protein binds to single binding sites in
two binding domains (each binding domain has $K$ binding sites),
leading to a binding  free-energy gain of $f_b$. A DNA loop leads to an
entropy loss $s$.}
\label{fig:model}
\end{figure}

In this paper, we assume  a more global point of view: May DNA loops
and looping proteins agglomerate  collectively to give rise to
transcriptional foci?  What are the thermodynamic ingredients leading
to such an agglomeration?  In this context, we model the DNA as a
string containing many {\it protein binding domains} (BD), each one
composed  of $K$ {\it binding sites} (BS).  In this work we consider
only bivalent DNA-binding proteins which are able to bind
simultaneously to two different BDs, introducing thus a DNA loop 
\footnote{Note that real DNA-binding proteins recognize specific
binding sites along the DNA. This specificity allows for the 
simultaneous agglomeration of transcription factories containing
specific transcription factors, and thus the attraction of
specific binding sites. The basic mechanism described here is,
however, not affected.}.
Fig.~\ref{fig:model} resumes the basic model ingredients.  We find
that this simple model leads to an effective attraction between DNA
loops and thus to the formation of protein agglomerates.

The role of multiple binding sites for a single loop has already been
studied by Vilar et al. \cite{Vilar,Saiz}, whereas multiple loops have
been considered by Hanke and Metzler \cite{Metzler} but for BDs with
only $K=1$ BS. We will show that only the  combination of both is able 
to introduce the desired emergence of protein agglomerates. This raises
an interesting question: given a concentration of looping proteins
and an entropy cost of bringing  two binding domains in close
vicinity, is it possible to get an agglomerate of binding  domains?
Besides being of interest to transcription factories the question is
actually very  general and interesting by itself when reformulated
differently: If we consider each BD to be a  monomer, then the problem
is equivalent to understanding the effect of introducing $L$ non nearest
neighbour links between $N$ monomers, with global constraints resulting
from the entropy loss due to DNA looping, and local constraints due to
the structure assumed for the BD.

In the following Sec.~\ref{sec:mechanism}, we first discuss the basic 
mechanism for protein agglomeration resulting from the combination of 
these ingredients. Later in this paper, in Sec.~\ref{sec:meanfield}, we 
introduce a mean-field model which can be mapped to a restricted 
random-graph ensemble. In Sec. \ref{sec:graph},  we solve 
it approximately generalizing a microscopic mean-field approach developed 
by Engel {\it et al.}~\cite{Engel}. In Sec.~\ref{sec:discussion}, 
we discuss the results of our mean-field  model in the context of factories 
and compare them with the known results for collapse of randomly linked 
polymers.

\section{The basic mechanism}
\label{sec:mechanism}

As shown in Fig.~\ref{fig:model},  there are two competing effects
related to DNA looping: First, the binding of a linking protein
introduces some free-energy difference $-f_b$ (for example in case of
lac operon $f_b$ is  of order 10-15 kcal/mol \cite{Vilar2}). The
second contribution comes from the fact that  each loop reduces the
conformational entropy of the DNA, thus a link leads to a total
free-energy  difference of $\Delta F = -f_b + T s$, with $T$ being the
temperature and $s$ being the entropy loss. In principle $s$  depends
on the length of the loop and on the DNA stiffness, cf.
\cite{Metzler}. For this qualitative argument we do not take care of
this  dependence and use the entropy loss of a typical-length loop.

\begin{figure}[htb]
\vspace{0cm}
\begin{center}
\includegraphics[width=0.6\columnwidth]{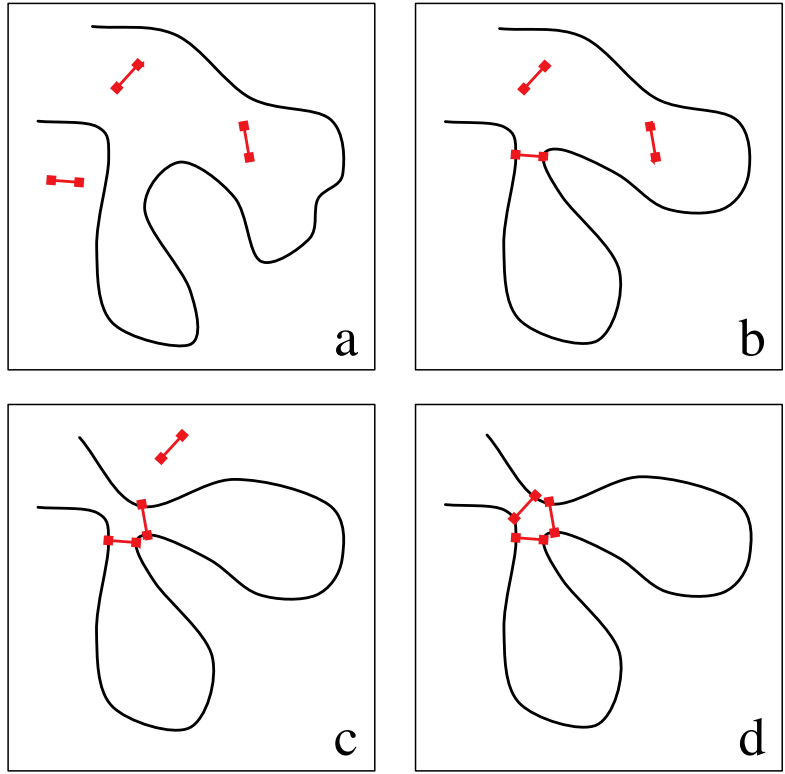}
\end{center}
\caption{Basic agglomeration mechanism:  (a) DNA is represented by a
string, binding proteins by linkers.  (b) Binding a protein to two BDs
leads to a gain of binding free energy,  but causes a loss in DNA
conformational entropy.  (c) The same happens, if a second loop is
introduced.  (d) Now, if one BD is common to the loops, a next protein
may bind to  the still unlinked BDs without major entropy losses. }
\label{fig:mechanism}
\end{figure}

Now, as shown in Fig.~\ref{fig:mechanism}, we introduce a second loop,
and the total free-energy difference to the unlooped configuration
becomes $\Delta F = -2 f_b + 2T s$. There are two possible cases for
the  relative positions of the two loops: First, the loops are
distant, and the binding of another linker protein has to introduce a
new loop. Second, loops share one BD. Then also the unconnected BDs of
the two loops may be linked, cf.~cases (c) and (d) in the figure.  In
this case, binding free energy is gained, but no new loop is
introduced, i.e., no  further entropy is lost.  We thus have a free
energy  $\Delta F = -3 f_b + 2T s$ which is lower than the one
achievable by distant loops. This mechanism introduces an {\it
effective attraction between binding domains of loops}: A cluster of
$n$ loops might be connected by $n(n+1)/2$ proteins, so the binding
free-energy is growing quadratically with the entropy loss. Note that
this picture is based on the simple observation of multiplicity of
protein binding sites  in a binding domain on DNA.

\section{A mean-field description via random protein-connection graphs}
\label{sec:meanfield}

To gain a first understanding of the action of this effective
attraction, we set up a mean-field model. The entropy loss $s$
due to the introduction of a loop is assumed to be independent of the
one-dimensional distance between two BDs measured along the DNA chain. 
Note that this approximation would be exact for monomers in a box.
 
On this level, BDs can be seen as {\it vertices} of a {\it
protein-connection graph}, and each bound protein between two such
vertices forms an {\it edge}. We assume $L$ proteins to be bound. The
entropy loss due to this linking depends on the component structure of
the graph: A connected component  (CC) of $n$ vertices contains $n-1$
loops. Denoting the number of CCs of $n$ vertices by $C(n)$, and
the  total vertex number by $N$, we find that the free-energy
difference with respect to the loop-free system is
\begin{eqnarray}
\Delta F &=& - L f_b + T s \sum_{n=1}^N (n-1) C(n)  \nonumber\\
&=& - L f_b - T C s +const.
\label{eq:DeltaF}
\end{eqnarray}
with $C=\sum_n C(n)$ being the total number of CCs.  This
free energy has two competing negative contributions. The first term
favors large $L$ by binding more proteins, and its ground state would
be the fully connected graph which has only one CC.  The second
contribution in (\ref{eq:DeltaF}) favors many components for positive
$s$. Its ground state is thus the empty  graph with each of the $N$
isolated vertices as a CC. The global behavior of  the model is given
by the balance of these two terms, and can be characterized by the
partition sum running over all graphs,
\begin{equation}
Z = \sum_{\rm graphs} \exp\{ L f_b/T + C s \}\ .
\label{eq:Z}
\end{equation}

We note that this partition function describes a modified random-graph
ensemble which depends only on the number of links and the number of
CCs. In fact, in usual diluted random graphs \cite{Erdos} each pair of
vertices is connected with some probability $0<p<1$, and left
unconnected  with $1-p$.  The probability of a specific graph with $L$
edges is then proportional to $[p/(1-p)]^L$, so it is  exponential in
the number of edges. If, further on, we reweight all graphs by some
factor $q^{C}$, we find that the graphs have a probability
corresponding to Eq.~(\ref{eq:Z}) by identifying  $p/(1-p) :=
e^{f_b/T}$ and $q:= e^s$. Further more, the sum over all graphs is
restricted by the connectivity constraint: At most $K$ proteins can be
bound to one BD, for $d$ bound proteins the distinguishable nature of
the BS inside the BD results in a  combinatorial factor
$K!/(K-d)!$. In the next section we give an analytical description 
of this problem for any $K$.

The main question of this work, i.e. the question if an agglomerate
exists or not, translates to the problem of graph percolation: Agglomeration
is equivalent to the existence of an extensively large connected
component in the graph, i.e. to the existence of a {\it giant component}.

\section{Analytical description of the graph ensemble} 
\label{sec:graph}

Before coming to the full description of the problem, i.e. to random 
graphs which have restricted degrees and are reweighted according to
their number of CCs, we concentrate a moment on the case $q=1$, i.e.
without considering the number of CCs. The basic idea is that the 
number of CCs can be introduced in a later moment considering large
deviations from typical $q=1$-graphs. The approach generalizes the
cavity-type calculation of \cite{Engel}, which is the special case
$K=\infty$. We will resort to this limit in order to check correctness
of our results.

\subsection{Graphs at $q=1$}

First, we describe the graphs without any constraint coming from
the number of CCs, i.e.~without entropy losses
in the DNA-looping model. In this case we have $s=0$, i.e. $q=1$.
The graphs have $N$ vertices (or BDs), each of them containing $K$ 
distinguishable BS (called stubs in the following) which allow vertices 
to have up to degree $K$. We describe graphs at the level of vertices 
by their symmetric adjacency matrix $\{J_{ij}\}$ with entries 1 whenever 
to vertices are connected via any two of their BS, and 0 else. The 
distinguishable nature of the binding sites is taken into account by a c
ombinatorial factor $K!/(K-d)!$ for any vertex of degree $d$. This factor 
counts the number of non-equivalent way the $d$ edges can be attached to 
the $K$ BS.

The statistical properties of a graph $G$ with adjacency matrix 
$\{J_{ij}\}$ can be characterized by its number of links
\begin{equation}
L(G) = \sum_{i<j} J_{ij}
\end{equation}
and its degree distribution given via the number $N_d$ of vertices
of degree $d$, 
\begin{equation}
N_d(G) = \sum_i \delta\left( d, \sum_j J_{ij} \right)\ ,\ \ \ d=0,...,K,
\end{equation}
with the notation $\delta(\cdot,\cdot)$ for the Kronecker symbol. 
Obviously these quantities are not independent. We have 
$\sum_d N_d = N$ and $\sum_d d N_d = 2L$ since links are counted
twice by adding up all degrees.
Without reweighting graphs by the number of CCs, the graph ensemble is
completely characterized by these quantities. In fact we write
\begin{eqnarray}
P(G|\gamma,K,N) &=& \frac 1{Z(\gamma,K,N)}\ \prod_{i<j}
\left[\left(1-\frac \gamma{K^2 N} \right) \delta(J_{ij},0)
+ \frac \gamma{K^2 N}  \delta(J_{ij},1) \right] 
\ \prod_{i=1}^N \left[ \frac{K!}{(K-\sum_j J_{ij})!} \,
\Theta( K-\sum_j J_{ij} ) \right]\nonumber\\
&=& \frac 1{Z(\gamma,K,N)}\ \left( \frac \gamma{K^2 N} \right)^{L(G)}
\ \left( 1-\frac \gamma{K^2 N} \right)^{{N\choose 2}-L(G)}\ 
\prod_{d=0}^K \left[ \frac{K!}{(K-d)!} \right]^{N_d(G)}\ ,
\label{eq:p_graph}
\end{eqnarray} 
where the last line already takes into account that degrees beyond $K$
are forbidden. In this notation, $\gamma$ acts as a chemical potential 
for links, and corresponds to the binding free energy in the protein
case. Note that the combination $\gamma/(K^2N)$ is chosen such 
that, for $K\to \infty$, we recover normal Erd\"os-Renyi random graphs
with average degree $\gamma$. In this limit, all our results here have
to coincide with the ones of \cite{Engel}.

From this microscopic description by the adjacency matrix of a graph,
we can go directly to a coarser description giving the probability of
a graph to have some degree distribution $N_d=p_d N$. We find
\begin{eqnarray}
P( N_0,...,N_K | \gamma,K,N ) &=& 
\frac 1{Z(\gamma,K,N)}\ \left( \frac \gamma{K^2 N} \right)^{L}
\ \left( 1-\frac \gamma{K^2 N} \right)^{{N\choose 2}-L}\ 
\prod_{d=0}^K \left[ \frac{K!}{(K-d)!} \right]^{N_d}
\nonumber\\
&& \times \frac{N!}{\prod_d N_d!}\ (2L-1)!!\
\prod_{d=0}^K \left[ \frac 1{d!} \right]^{N_d}\ .
\end{eqnarray} 
This equation is obtained by multiplying the probability of a single
of these graphs by the number of possible realizations of the degree
sequence. The factors are, in the order of appearance, the number of
ways to assign degrees to the $N$ vertices, the number of 
possibilities to wire a set of vertices with given degrees,
and a correction of overcounting due to the (at this point) 
indistinguishable occupied BSs. Note that the factor 
$(2L-1)!! = (2L)!/2^{L}/L!$ can be easily understood in terms of 
generating the graph: Once degrees are assigned, each vertex of degree
$d$ is assigned $d$ stubs (or half-edges). These are brought into a
random permutation, and the first stub becomes connected to the second 
one, the third to the fourth etc. This procedure overcounts graphs:
A factor $2^{-L}$ accounts for possible permutation of the two stubs
inside each of the $L$ links, the factor $1/L!$ for permutations of
entire links. This procedure does not forbid double links or self-loops,
but these are rare and therefore do not influence the global statistical
features of the graph ensemble. Using Stirlings formula 
$\ln N! = N\ln N - N + o(N)$ we can rewrite this as
\begin{eqnarray}
P( N_0,...,N_K | \gamma,K,N ) &=& \frac 1{Z(\gamma,K,N)}
\nonumber\\ && \times
\exp\left\{ N \left[ \ell \ln \left(\frac \gamma{K^2 N}\right)
-\frac{\gamma}{2K} + \sum_d p_d \ln {K \choose d}
-\sum_d p_d \ln p_d + \ell \left( \ln[2\ell N] -1 \right)
\right] +o(N)\right\} \nonumber\\
&=& \frac 1{Z(\gamma,K,N)} \ 
\exp\left\{ N \left[ - \sum_d p_d \ln\left( {K\choose d}^{-1} p_d \right)
+\ell \ln \left( \frac{2\gamma\ell}K \right) - \ell
\right] + o(N)\right\}\ ,
\label{eq:p_degrees}
\end{eqnarray} 
where we have used $\ell = L/N = \frac 12 \sum_d d p_d$. The typical degree 
distribution, which is realized in this ensemble with probability 
tending to one in the thermodynamic limit, can be evaluated by the maximum 
of this expression. Deriving the exponent by $p_d$, and using a Lagrange
multiplier to ensure normalization of the degree dsitribution, we
arrive directly at (overbars always denote typical values)
\begin{equation}
\overline p_d = {K\choose d} \frac{z^d}{(1+z)^K}\ ,\ \ \ \ 
z = \sqrt{ \frac {2\gamma\overline \ell}{K^2} }\ .
\label{eq:degrees}
\end{equation}
With the average degree determined from this distribution,
\begin{equation}
\overline d = 2\overline \ell = K \frac z{1+z}\ ,
\end{equation}
we arrive at two self-consistent equations for $z$ and $\overline d$ which
are solved by (a second non-physical solution is not shown)
\begin{eqnarray}
z &=& \frac 12 \left( \sqrt{ 1+4 \frac \gamma K } -1 \right) \nonumber\\
\frac {\overline d}K &=& 
\frac{\sqrt{1+ 4 \frac \gamma K } -1}{\sqrt{ 1+4 \frac \gamma K } +1}\ .
\label{eq:zd}
\end{eqnarray}
Note that, for $K\to\infty$, the degree distribution (\ref{eq:degrees})
tends as expected to the Poissonian law $e^{-\gamma} \gamma^d/d!$.
These results allow us to express also the dominant contribution to
the partition function $Z(\gamma,K,N)$ by the exponent in 
Eq.~(\ref{eq:p_degrees}) evaluated at the saddle point:
\begin{equation}
\frac 1{KN} \ln Z(\gamma,K,N) = - \frac {\overline d} K \ln \frac {\overline d} K
- \left(1-\frac {\overline d} K\right) \ln \left( 1- \frac {\overline d} K \right) 
+ \frac {\overline d}{2K} \ln \left( \frac \gamma K \frac {\overline d} K \right) 
- \frac {\overline d}{2K} +o(1)\ .
\label{eq:z}
\end{equation}

\subsection{The case $q\neq 1$}

Now we have all results being important for analyzing the full model including
entropy losses. In difference to the case discussed so far, the full graph 
ensemble includes a weight $q^{C(G)}$ depending on the number $C(G)$ of
connected components of graph $G$. We therefore consider the 
modified ensemble
\begin{equation}
\label{eq:Pq}
P(G|\gamma,q,K,N) = \frac 1{Z(\gamma,q,K,N)} P(G|\gamma,q,N) q^{C(G)},
\end{equation}
where the normalizing partition function is given as
\begin{equation}
Z(\gamma,q,K,N) = \sum_G P(G|\gamma,q,N) q^{C(G)}\ .
\end{equation}
Note that this partition function fulfills $Z(\gamma,q\!=\!1,K,N)=1$ due
to the use of the normalized distribution $P(G|\gamma,q,N)$ in
Eq.~(\ref{eq:Pq}).

Compared to the previous section, this graph ensemble is hard to
handle: The distribution depends on the global quantity $C(G)$ which
cannot calculated as easily as the degree distribution, which was 
sufficient to characterize the graph ensemble at $q=1$. To get 
information, we modify the cavity-type approach of \cite{Engel}.
There, information on the graph ensemble with 
$C$-dependent weight (but without degree constraint, i.e. the case
$q\neq 1$ and $K\to \infty$) was obtained via a simple and intuitive
idea: If we add a new vertex to a typical graph of size $N$, and the 
degree of this vertex is randomly selected according to the typical 
degree distribution of size-$N$ graphs, the new graph is basically 
equivalent to a typical graph of size $N+1$. Unfortunately, this 
argument holds only approximately in our case. The added vertex does not
become a typical one of the enlarged graph. However, since it holds for 
$K\to\infty$ and for $K=1$, so we expect it to be rather precise also 
for intermediate-large $K$.

Assume therefore that we add a vertex to a graph $G$ of $N$ vertices.
The degree $d$ of this vertex is drawn randomly from 
$\overline p_d = {K \choose d} z^d / (1+z)^K$
with $z$ given by Eq.~(\ref{eq:zd}) (graph ensemble at $q=1$). 
Now, the number of components changes according to
\begin{equation}
P(C|\gamma,K,N+1) = \sum_{\Delta C} \tilde D(\Delta C) 
P(C+\Delta C|\gamma,K,N)\ .
\label{eq:addvertex}
\end{equation}
The kernel $\tilde D(\Delta C)$ can be decomposed into various
contributions according to the degree $d$ of the new vertex, and
the number $d_0$ of links this new vertex makes with the giant
component of $G$. The change $\Delta C$ of the number of CCs also
depends on these two numbers. For positive $d_0>0$, we add
$d-d_0$ small components to the giant one, i.e. we have
$\Delta C=d-d_0$. For $d_0=0$, we unify $d$ small components to
a single one, and we have $\Delta C = d-1$. The kernel therefore
results in
\begin{eqnarray}
\tilde D(\Delta C) &=& \sum_{K\leq d\leq d_0 \leq 0} 
\tilde D(\Delta C,d,d_0)\ , \nonumber\\
\tilde D(\Delta C,d,d_0) &=& \alpha \overline p_d {d \choose d_0}
\pi^{d_0} (1-\pi)^{d-d_0} \delta( \Delta C, d-d_0-\delta(d_0,0))\ ,
\label{eq:Dtilde}
\end{eqnarray}
with $\pi$ denoting the probability of selecting an end-vertex inside
the giant component. Due to the special definition of our ensemble, 
where we do not select directly vertices but free BS associated to a
vertex, the number $\pi$ equals therefore the fraction of all 
{\it free BS} being inside the giant component. It has a simple relation 
to the fraction $\nu$ of {\it vertices} belonging to the giant component:
\begin{equation}
\pi = \frac {K-\overline d_{in} }{K-\overline d} \nu\ .
\label{eq:pinu}
\end{equation}
Here $\overline d_{in}$ denotes the average degree inside the giant
component,  $\overline d$ the one of the full graph.

The reweighting factor $\alpha$ can be calculated exactly in the
case $K\to\infty$, cf.~\cite{Engel}. Its precise value is not of
interest in our discussion. If we multiply Eq.~(\ref{eq:addvertex}) 
by $q^C$ and sum over $C$, we obtain
\begin{eqnarray}
Z(\gamma,q,K,N+1) &=& \sum_{\delta C} \tilde D(\Delta C) q^{-\delta C}
Z(\gamma,q,K,N) \nonumber\\
&=& \zeta(\gamma,q,K)\ Z(\gamma,q,K,N)\ .
\end{eqnarray}
The logarithm of $\zeta(\gamma,q,K)$ can be interpreted as a 
free-energy shift in the graph ensemble due to adding a new vertex.
Using Eq.~(\ref{eq:Dtilde}) it can be calculated right away, resulting
in 
\begin{equation}
\zeta(\gamma,q,K) = \alpha \left[ \frac{q+z(1-\overline\pi)}{q(1+z)}
\right]^K \left[ q-1+\left\{ 1+ \frac{qz\overline\pi}{q+z(1-\overline\pi)}
\right\}^K \right]\ .
\end{equation}
Note that due to the concentration of intensive quantities to their
typical values in the summation over all graphs, we have replaced
$\pi$ by its saddle-point value $\overline \pi$.

The decomposition of  $\tilde D(\Delta C)$ helps to get more detailled 
insight into the graph structure. In fact, the quantity
\begin{eqnarray}
\zeta(d,d_0|\gamma,q,K) &=& \sum_{\Delta C} \tilde D(\Delta C,d,d_0)\ 
q^{-\delta C} \nonumber\\
&=& \alpha \overline p_d  {d \choose d_0}
\overline\pi^{d_0} (1-\overline\pi)^{d-d_0} q^{-d+d_0+\delta(d_0,0)}
\end{eqnarray}
describes an effective single-vertex Boltzmann factor, and 
single-vertex quantities as the probability of belonging to the
giant component or the degree distribution can be derived from it.

To start with the giant component size, we remind that for all 
$d_0>0$ the newly added vertex becomes connected to the giant component,
and thus is part of it in the $(N+1)$-vertex graph. Therefore the 
fraction of vertices {\it not} belonging to the giant component
can be written as
\begin{eqnarray}
1-\overline\nu &=& \frac 1{\zeta(\gamma,q,K)} \sum_{d=0}^K
\zeta(d,d_0=0|\gamma,q,K) \nonumber\\
&=& \frac q { q-1+\left\{ 1+ \frac{qz\overline\pi}{q+z(1-\overline\pi)}
\right\}^K}  \ .
\label{eq:gc}
\end{eqnarray}

The degree distribution of the graph results in
\begin{eqnarray}
P(d|\gamma,q,K) &=& \frac 1{\zeta(\gamma,q,K)} \sum_{d_0=0}^d
\zeta(d,d_0|\gamma,q,K) \nonumber\\
&=& {K\choose d}  \ \frac
{ \left[ \frac {z(1-\overline\pi)} q \right]^d \left[
q-1 +
\left\{ 1+q \frac{\overline\pi}{1-\overline \pi} \right\}^d
\right]}
{\left[ \frac{q+z(1-\overline\pi)}{q}
\right]^K \left[ q-1+\left\{ 1+ \frac{qz\overline\pi}{q+z(1-\overline\pi)}
\right\}^K \right]}\ .
\end{eqnarray}
For $q\neq 1$, this distribution deviates from a simple binomial
distribution. The average vertex degree follows immediately,
\begin{eqnarray}
\frac {\overline d}K &=& z\ \frac
{ (1-\overline\pi)(q-1) + [1+\overline\pi(q-1)]
\left\{ 1+ \frac{qz\overline\pi}{q+z(1-\overline\pi)}
\right\}^{K-1} }
{ [q+z(1-\overline\pi)] 
\left[ q-1+\left\{ 1+ \frac{qz\overline\pi}{q+z(1-\overline\pi)}
\right\}^K \right]}
\nonumber\\ &=&
\frac {z(1-\overline\nu)(1-\overline\pi)}{q  [q+z(1-\overline\pi)] }
\left[q-1 + 
\left\{1+ \frac{q\overline\pi}{1-\overline\pi} \right\}
\left\{ 1+ \frac{q\overline\nu}{1-\overline\nu} 
\right\}^{\frac{K-1}K} \right]\ ,
\label{eq:l_vertex}
\end{eqnarray}
where we have used Eq.~(\ref{eq:gc}) to simplify the expression in the
second line. The degree distribution inside (resp. outside) the giant
component can be obtained by restricting sums to $d_0>0$ (resp.
$d_0=0$),
\begin{eqnarray}
P_{\rm in}(d|\gamma,q,K) &=& \frac
{\sum_{d_0=1}^d \zeta(d,d_0|\gamma,q,K)}
{\sum_{d=1}^\infty\sum_{d_0=1}^d \zeta(d,d_0|\gamma,q,K)}\ ,
\nonumber\\
P_{\rm out}(d|\gamma,q,K) &=& \frac
{\zeta(d,d_0=0|\gamma,q,K)}
{\sum_{d=0}^\infty \zeta(d,d_0=0|\gamma,q,K)}\ .
\end{eqnarray}
For the outside average degree we obtain the particularily simple
result
\begin{equation}
\overline d_{\rm out} = K \frac{z(1-\overline\pi)}{q+z(1-\overline\pi)}
\ .
\label{eq:dout_vertex}
\end{equation}
For the inside average degree, we use the fact that the total average
degree is the weighted average of the inside and outside degrees,
and we find
\begin{equation}
\overline d_{\rm in} = \frac {\overline d - \overline d_{\rm out}
(1-\overline\nu)}{\overline\nu}\ .
\label{eq:din_vertex}
\end{equation}

\subsection{The phase diagram}

We have now enough equations to determine self-consistently the phase 
diagram. Putting together
Eqs.~(\ref{eq:pinu},\ref{eq:gc},\ref{eq:l_vertex},\ref{eq:dout_vertex})
and (\ref{eq:din_vertex}), and eliminating directly the expression for
$\overline d_{\rm in}$, we find a closed set of three equations
\begin{eqnarray}
\label{eq:sp}
\overline \pi &=& \frac 1{K-2\overline \ell} \left[ K \overline \nu -
2\overline \ell +K \frac{z(1-\overline \pi)(1-\overline \nu)}
{q+z(1-\overline \pi)} \right]\nonumber\\
\overline \ell &=& \frac K2\
\frac {z(1-\overline\nu)(1-\overline\pi)}{q  [q+z(1-\overline\pi)] }
\left[q-1 + 
\left\{1+ \frac{q\overline\pi}{1-\overline\pi} \right\}
\left\{ 1+ \frac{q\overline\nu}{1-\overline\nu} 
\right\}^{\frac{K-1}K} \right]\\
\overline\nu &=& 1-
\frac q { q-1+\left\{ 1+ \frac{qz\overline\pi}{q+z(1-\overline\pi)}
\right\}^K}\ .
\nonumber
\end{eqnarray}
We had introduced the model in function of the parameter $\gamma$ which
can be understood as a chemical potential coupled to the number of edges
in the graph. Since this parameter has no very obvious interpretation due
to its interaction with the degree-constraints, we prefer to use its
conjugate quantity, the link density $\overline \ell$ as a control
parameter. In this sense, for given $K$, the phase diagram is spanned by
$q$ and $\overline \ell$, and Eqs.~(\ref{eq:sp}) allow to determine
the unknown quantities $\overline \nu, \overline \pi$ and $\gamma$.

Eqs.~(\ref{eq:sp}) always have the solution
$\{\overline \nu,\overline \pi,\gamma\}=\{0,0,2\overline\ell qK/(K-2\overline\ell)\}$. 
It  corresponds to a phase without any extensive CC, i.e. to a non-agglomerated 
phase.  For large enough $\overline \ell$ and $q=e^s$, also other solutions exist. 
To see this, we expand Eqs.~(\ref{eq:sp}) up to second order in $\overline \nu$ and 
$\overline \pi$, and find in particular
\begin{equation}
\overline \pi=-\frac{K[2\overline \ell(K-1)-K]}{2\overline \ell^2 (K-1)[2+K(q-2)]}
\label{eq:pi_cont}
\end{equation}
which implies a continuous transition to a non-trivial solution at
\begin{equation}
\overline \ell_c= \frac K{2(K-1)},~~\forall~q<q_c = 2- \frac 2K
\end{equation}
Note that, for $K\to\infty$ and $q=1$, this result reproduces the known 
percolation result in Erd\"os-R\'enyi random graphs. For $q<q_c$, for 
all $K$, Eq.~(\ref{eq:pi_cont}) implies that $\overline \pi \sim (\overline \ell
- \overline \ell_c)$ near 
the transition point. At $q=q_c$, we can expand Eqs.~(\ref{eq:sp}) up to
third order in $\overline \nu$ and $\overline \pi$.  We find that there is 
a percolating point at same value of $\overline \ell_c$ as for $q<q_c$, but with  
$\overline \pi \sim (\overline \ell - \overline \ell_c)^{1/2}$. 
Note that this transition exists for all $K>2$, at $K=2$ itself the transition 
point would be $\overline\ell_c=1$ which equals the highest possible  
degree in this graph (due to $2\overline \ell\leq K$).

\begin{figure}
\begin{center}
\includegraphics[width=0.6\columnwidth]{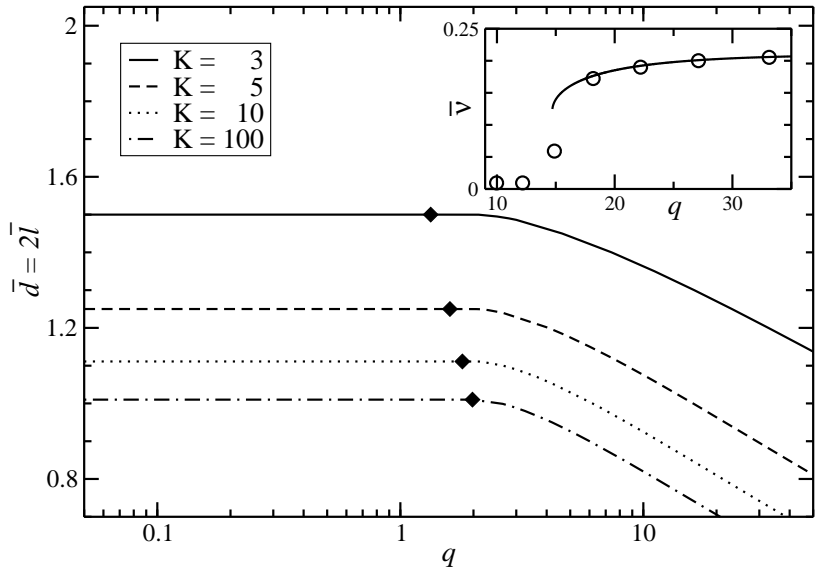}
\end{center}
\caption{ Phase diagram
for protein agglomeration in mean-field description for
$K=3,5,10,100$: Below the line, no extensive CC exists. Above, a
finite fraction of all links and vertices is collected in the largest
CC. The transition is continuous on the left,  discontinuous on the
right side of the diamonds.  Inset: Fraction of vertices collected in
the largest component as function of $q=e^s$, for $\overline \ell = 0.2$ 
and $K=20$. The full line is the analytical result of Eqs.~(\ref{eq:sp}),  
the symbols show results of MC simulations for $N=5000$,  each symbol is
an average over 900 independent equilibrium configurations.}
\label{fig:phasediagram}
\end{figure}

For $q>q_c$, Eq.~(\ref{eq:pi_cont}) does not make sense. We find  a
{\it discontinuous} transition at smaller $\overline \ell_c(q,K)$ which has 
to be determined from Eqs.~(\ref{eq:sp}) via the spinodal point; the 
transition point can be
obtained with good precision using symbolic manipulation software
like Mathematica \cite{Mathematica}. In this case, the largest
component jumps from a non-extensive size to a finite fraction of the
full system. In Fig.~\ref{fig:phasediagram}, the phase diagram for
various values of $K$ is given. It is found to be qualitatively
similar for all $K\geq 3$, but agglomeration is favored for
higher-order BDs at same number of links. In the inset of the figure 
we show also the discontinuous nature of the transition: At given number
of links the parameter $q$ is changed, and the size of the largest CC
is recorded. We find an excellent agreement between MC simulations of 
random graphs using Metropolis type rewiring steps, and  the analytical
results obtained  from Eqs.~(\ref{eq:sp}). This illustrates the quality of
approximation done in the analytical approach under vertex addition.

It is very interesting that the phase transition appears at smaller
$\overline \ell$ for higher entropy losses $s=\ln q$. The reason is that an
increased $s$ leads to a compaction of the giant component, where links can 
be added without loosing further CCs. Therefore, even if the
transition appears at lower global average degree, the  average degree
inside the largest CC always exceeds two. Again, this fact illustrates 
why $K>2$ is essential for agglomeration.

\begin{figure}[htb]
\begin{center}
\vspace{1 cm}
\includegraphics[width=0.6\columnwidth]{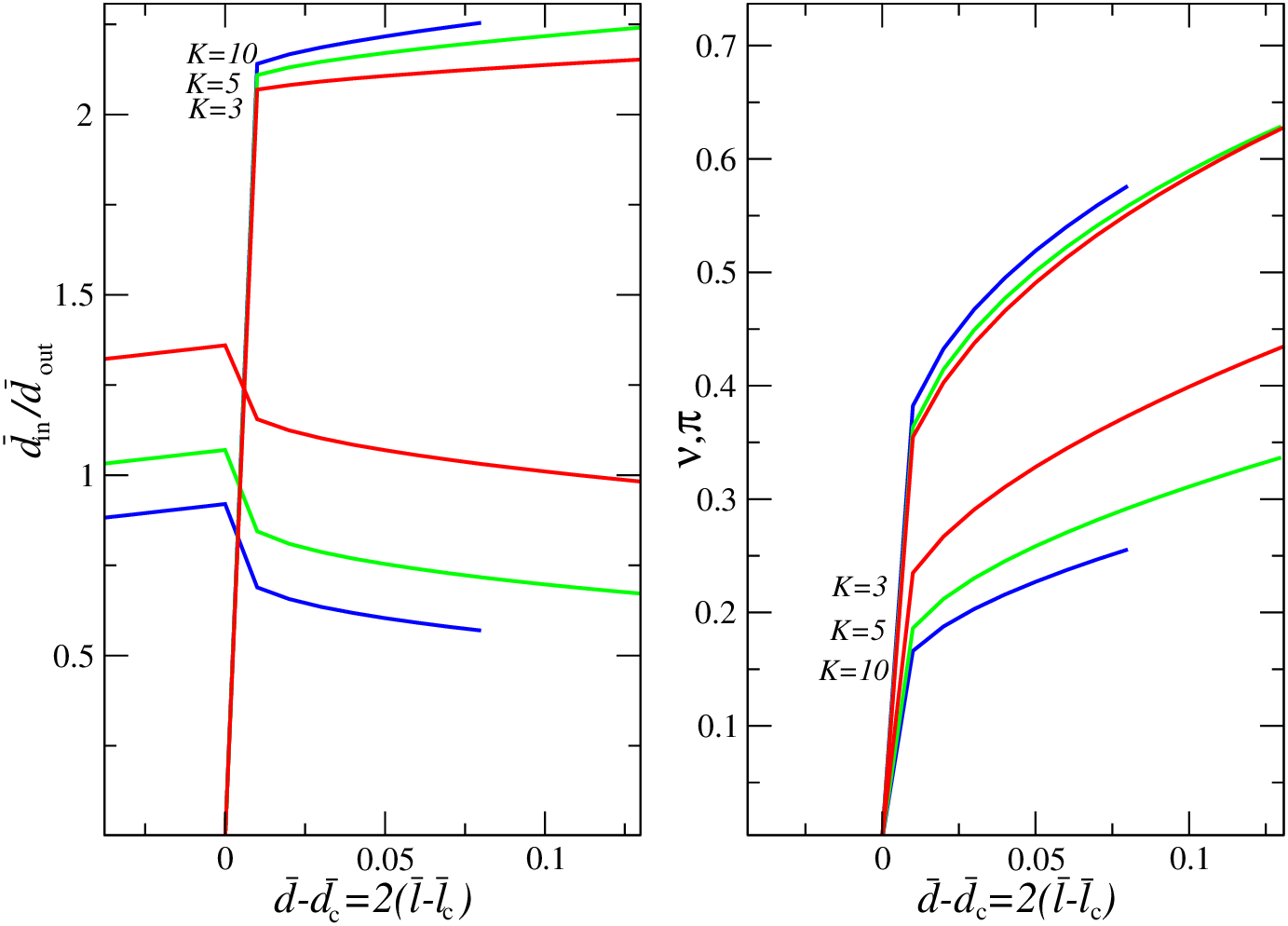}\hspace{1cm}
\end{center}
\caption{Left panel: Average degrees $\overline d_{\rm in}$ inside (upper curves) and  
$\overline d_{\rm out}$ outside the giant component (lower curves), as a 
function of the total link density $\overline \ell$, for values $K=3,5,10$ and 
fixed $q=10$.
Right panel: Fraction $\overline\nu$ of vertices (lower curves) and fraction of
links (upper curve) for the same parameters as in the left panel. }
\label{fig:din_dout}
\end{figure}

The left panel of Fig.~\ref{fig:din_dout} shows the plot of the average degrees
inside and outside the giant component for $q=10$, and for different values 
of $K$. For all  $K\ge 3$, both $\overline d_{\rm in}$ and $\overline d_{\rm out}$ 
show clearly a discontinous jump at a critical $K$-dependent value of 
$\overline \ell$. On one hand, $\overline d_{\rm in}$ always
jumps to a value slightly above two being necessary for the largest component to
be connected. The degree outsite the giant component, on the other hand, 
jumps to values which become smaller and smaller with increasing $K$.
In the right panel of Fig.~\ref{fig:din_dout}, we also show the fraction
of vertices resp. edges being in the giant component. For higher $K$ values,
the fraction of vertices becomes smaller at the transition point, whereas
the fraction of links becomes larger. This illustrates that, for larger $K$,
the agglomerate at the threshold becomes smaller but more dense.

\begin{figure}[htb]
\begin{center}
\includegraphics[width=0.6\columnwidth]{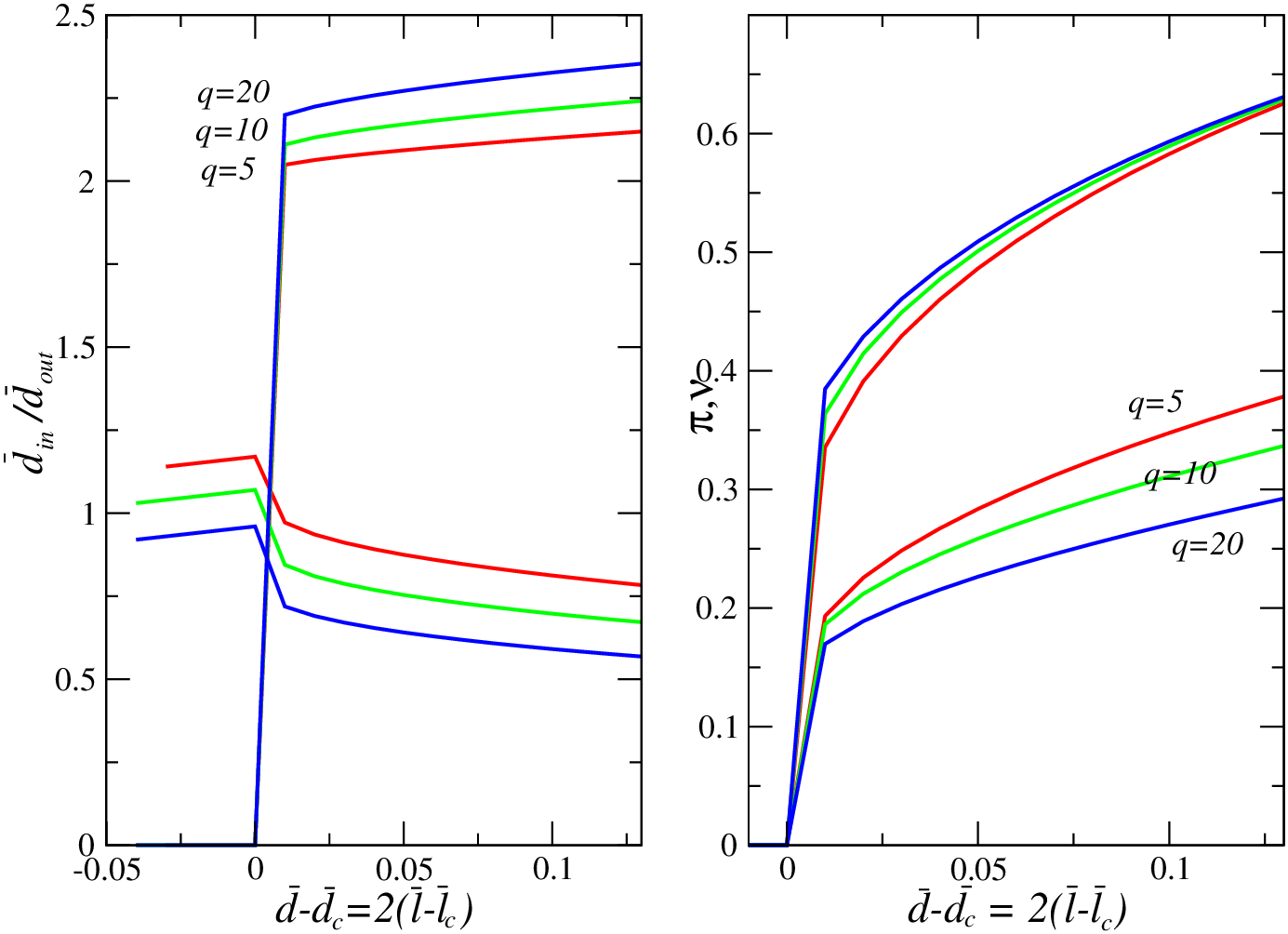}
\end{center}
\caption{Left panel: Average degrees $\overline d_{\rm in}$ inside (upper curves) and  
$\overline d_{\rm out}$ outside the giant component (lower curves), as a 
function of the shifted link density $\overline \ell - \overline\ell_c$, for fixed 
$K=5$ and diverse values of $q=5,10,20$.
Right panel: Fraction $\overline\nu$ of vertices (lower curves) and fraction of
links (upper curve) for the same parameters as in the left panel.}
\label{fig:K5}
\end{figure}

Similarly, it is illustrative to look at the properties of the giant component 
for various values of of $q$. Fig.~\ref{fig:K5} shows the plots for $K=5$ at 
$q=5,10$ and $20$. We see that, as $q$ increases, the fraction of vertices inside 
the giant component goes down (right panel), but the fraction of edges goes up, 
implying that the agglomerate becomes more and more compact with increasing $q$. 
Also the difference between the indegree and outdegree goes up as $q$ increases 
(left panel).

\section{Discussion} 
\label{sec:discussion}

The aim of the present paper is to present a minimal model which, on
the basis of a thermodynamic approach to DNA-protein interactions, is
able to show protein agglomeration. In this sense, it can serve as a
minimal model for the mechanism behind the formation of transcription
factories, which are observed in transcriptionally active cells. In
our paper we show that two ingredients are sufficient: DNA-looping
proteins which are able to bind  simultaneously to two -- also distant
-- protein binding sites on the DNA, and binding domains on the DNA
which contain, on average, more than two binding sites each. In this
case, the competition between free-energy gain by protein binding and
entropy loss by DNA looping is found to lead to an effective
attraction between DNA loops. As a consequence, binding domains and
proteins agglomerate collectively.

In its minimal character, the model might miss some important
properties of the biological system. As an example, we consider the
number of doubly-bound  proteins as one important control parameter,
whereas the relevant parameter should be the total number of proteins
- which would also include free and   singly bound molecules. Using
biologically reasonable parameters for the binding  affinities
(ca. 5-15 kcal/mol), we find in simulations that basically none of the
proteins stay free and a large majority is doubly bound in the
phase-transition  region. It would be interesting if we could extend
our analytical model accordingly.

Further more, our model did not consider the specificity of interaction
between DNA-binding proteins and their binding sites. This specificty
may result in the simultaneous agglomeration of various specific
transcription factories, which actually is an important ingredient
to \cite{Cook,Kepes}. Including this possibility into our model would
be another interesting generalization, but it would not effect the
very basic agglomeration mechanism.

Our model of agglomeration is similar to models of randomly linked
polymers studied  in the literature. Previous studies mostly
considered a Gaussian chain with randomly placed links using
variational  \cite{Bryngelson} and numerical methods
\cite{Kantor}. Bryngelson et al \cite{Bryngelson} in their study based
on variational approach and scaling arguments tried to argue that for
a Gaussian chain when the links are soft,  there is always a
transition. They also argued that it is a continuous transition that
occurs above a threshold which is a product of the density of links
and logarithm of average length of the loop. This result implied that
for some arbitrary polymer,  it is possible for transition to occur at
vanishing density of links  ($\sim 1/\ln N$). This result was
countered by Kantor et al \cite{Kantor}. Based on scaling arguments
they  argued that number of links $M$ necessary for a percolating
collapsed phase to exist scale as  $N^{\phi}$, with $\phi=1-1/d \nu$,
where $\nu$ is the exponent which describes the shape of the
polymer\footnote{this is different from the definition of $\nu$ used
in the rest of the paper}  (radius of gyration $R_g \sim
N^{\nu}$). Also, for the probability of looping $p(l) \sim
A/l^{\alpha}$,  with $1<\alpha<2$, it was shown by Schulman and Newman
\cite{Schulman} that for $M<N/2$ no infinite percolating cluster
exists.  For $M\ge N/2$, percolation may or may not occur depending on
the value of $A$ and $\alpha$.

Our solution correspond to the case where we have ignored length
dependence of the entropy cost of the  loop ($\alpha=0$) and we find
that there would always be a transition from extended phase to
collapsed phase, though  the nature of transition depends on the
entropy cost ($s=-\ln A$) of the loop. Most surprisingly, larger  the
cost of looping, smaller is the concentration at which the transition
happens. Also it is no longer a  second order percolation like
transition, but a first order transition for low concentration of
links.

\begin{figure}
\begin{center}
\includegraphics[width=0.85\columnwidth]{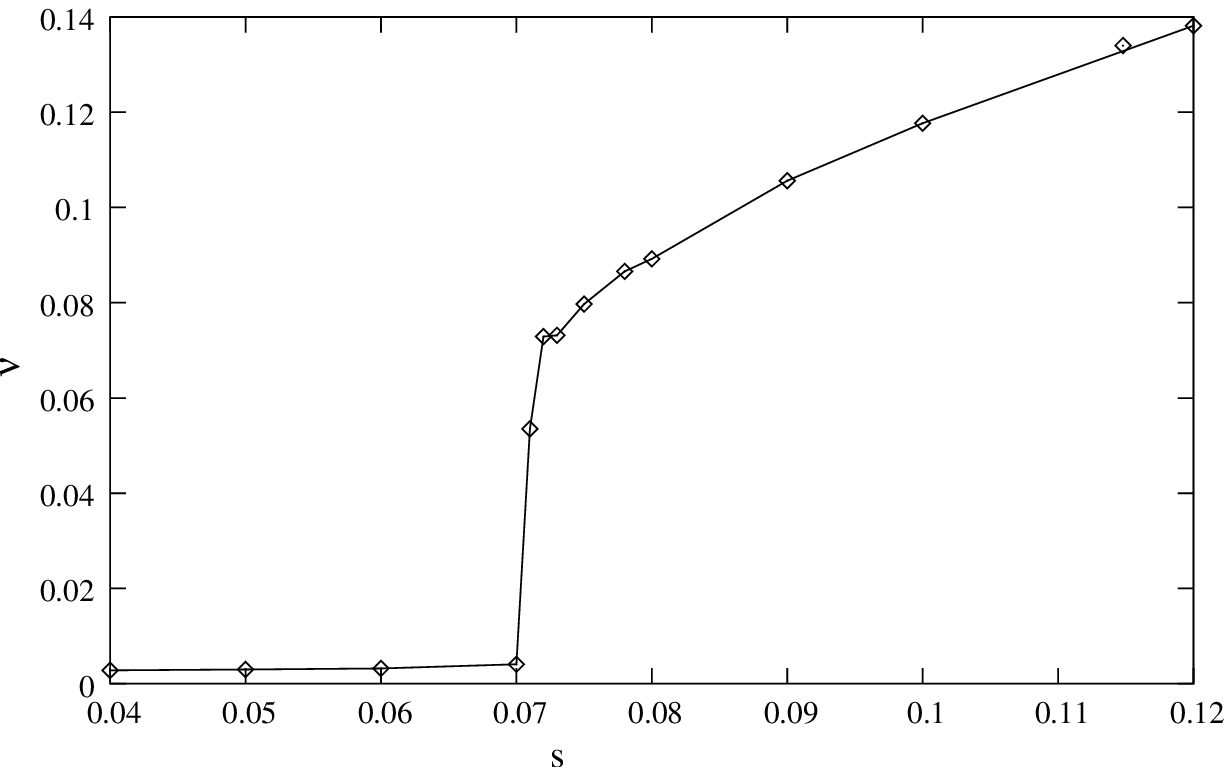}
\end{center}
\caption{We consider the contribution of entropy loss to the change in
free energy for a loop  between two BDs to be  equal to $s(-\ln
l_0+\frac{3}{2} \ln l)$. The figure plots the fraction of  sites in
the giant component for $K \rightarrow \infty$ as a function of $s$
for a chain with  $N=5000$ binding domains and $1000$ transcription
factors.}
\label{fig:lendep}
\end{figure}

We did simulations for a Gaussian chain  in three dimensions: In our 
mean-field model we considered entropy losses  to be independent of 
distance between the BDs measured along the DNA. The distance dependence 
of entropy in vivo is complicated. If we assume that the unlinked DNA
behaves on long scales like a Gaussian chain, the entropy loss is
monotonously increasing in the the loop length and scales as $q(l) =
e^{s(l)} \sim (l/l_0)^{3/2}$, where $l_0$ is the minimum distance
between two ends of a loop. If we now look to a connected component of
the $n$ vertices  $\{i_1,...i_n\}$ with $i_m<i_{m+1}$ for all $1\leq m
<n$, the entropy loss is given by $ s(i_1,...i_n) = \sum_{m=1}^{n-1}
s( i_{m+1}-i_m  )\, $.  In our simulations, we find that there is
still a discontinuous transition which  depends on the choice of $l_0$
(see Fig. \ref{fig:lendep}). Since longer loops are suppressed
compared to shorter ones, one could expect CC to be more localized in
one-dimensional distance along the DNA. This would correspond to
Cook's picture where DNA loops around a factory form a kind  of
rosetta, before DNA goes to the next factory. The logarithmic entropy
dependence taken into account in our simulations is not sufficient for
such a localization.

Based on these simulations, we suggest that even when we take length
dependence into account, there is a  possibility of first order
transition to the agglomerate, at small density of links for high
entropy cost. The reason this could be possible would be because of
the larger (exponential) contribution of the distribution of  links to
the entropy in comparision to the lograthimic dependence of entropy on
length.

We have ignored the interaction between loops. It is not clear how
important that could be. For example,  in the case of DNA denaturation
\cite{Poland}, exact results which ignore interaction between loops
predict a continuous transition in all dimensions less than
four. Whereas using scaling arguments and taking interaction of loop
with rest of the chain into account, Kafri et al \cite{Kafri}  showed
that the transition becomes first order in $d=2$.

The present work can be extended into various directions. First, from
biological point of view  it would be interesting to go to more
realistic modeling schemes (like worm-like chains for the DNA
molecule) and to check  the proposed picture. Such a simulation would
also allow to introduce biologically realistic parameters for protein
binding affinities and entropy losses, and to locate such a realistic
setting in the simplified mean-field phase diagram. However, current
simulations are concentrated  to a single loop \cite{Bon,Toan}, so
this task seems to pose a considerable  numerical challenge. It would
be interesting to see if self exclusion, depletion effects due to
macromolecular crowding or restricted volume would lead to a  spatial
localization of the agglomerate. A second direction could be the
inclusion of diverse looping proteins with specific binding sites on
the DNA to see whether equilibrium thermodynamics can drive the
creation of transcription factor specific spatial foci. Further, it
would be interesting to see  perhaps with the help of simulations for
Gaussian and self-avoiding chains if one can see a regime of
discontinuous transition from extended to collapsed phase.

{\it Acknowledgment} --- We warmly thank F. K\'ep\`es and O.  Martin
for many helpful discussions. We are also thankful to H. Orland for
pointing out the reference  \cite{Bryngelson} to us. This work was
support by the EC via the STREP GENNETEC (``Genetic networks:
emergence  and complexity'').



\end{document}